\documentclass[aps,prb,twocolumn,groupedaddress,showpacs,floatfix]{revtex4}
\usepackage{epsfig}
\begin{document}
%

\title{{\it Ab initio} investigation of the groundstate, magnetic,
  electronic, and optical properties of polyyne and cumulene
  prototypes}

\author{Carlo Motta}
\author{Marco Cazzaniga}
\author{Andrea Bordoni}
\author{Katalin Ga\'al-Nagy}
\email{katalin.gaal-nagy@physik.uni-r.de}
\affiliation{Universit\`a degli Studi di Milano, Dipartimento di
  Fisica, CNISM-CNR-INFM, via Celoria 16, I-20133 Milano, Italy}

\date{\today}
%
\begin{abstract}
  We have investigated polyyne and cumulene prototypes based on the
  density-functional theory. Our independent-particle spectra show
  that the various carbynes can be distinguished by optical properties
  comparing the low-energy spectral structure as well as using very
  general considerations. The latter conclusion is supported by
  results based on the random-phase approximation including
  local-field effects.
\end{abstract}
\maketitle

%
%
\section{Introduction}\label{Intro}

Carbon-based materials possess many potential ability for future
technologies.\cite{1}\cite{2}\cite{3} Carbon atoms (C) are known to
have three bonding states resulting from hybridization of the atomic
orbitals: ${sp}^{3}$ (diamond), ${sp}^{2}$ (graphite and fullerenes),
and $sp$. While the importance of ${sp}^{2}$ structure is by now
confirmed, a lot of interest is growing around $sp$ carbon allotropes,
named carbynes.

The formation of $sp$ carbon chains is expected in the initial stage
of small carbon cluster formation, on the road towards the fullerene
structure, when the ${sp}^{2}$ phase is energetically less
favorable.\cite{Lu05} Furthermore, the interpretation of some IR bands
from the interstellar dust could be related to the presence of
$sp$-coordinated carbon.\cite{Gu2008} Carbon chains are also
interesting systems because they can be found connecting different
graphene fragments.\cite{Rav09,Jin2009} Experimental work has been
recently devoted to the production and investigation of
$sp$-hybridized carbon
structures.\cite{Lu05,Mag2003,6,Rav2002,Sce2002,Jin2009} Furthermore,
pure-carbon solid containing carbynoid structures have been produced
by cluster beam deposition in the last years.\cite{6} Carbynes survive
landing because they are stabilized and protected by the ${sp}^{2}$
network.

Open carbon (C) chains of any length must be terminated by molecular
complexes to ensure stability, of course. In principle, linear carbon
chains can be obtained with two different terminations: $sp^3$
termination, resulting in carbon atoms linked by alternated single and
triple bonds (polyyne) and so with alternating bond length, and $sp^2$
termination, resulting in double bonds (cumulene).\cite{H04}
Experimentally polyynes are found to be more stable than
cumulenes.\cite{Gu2008} In this work, we have chosen hydrogen (H)
terminations. The choice of saturating the chains with one or two
hydrogen atoms reflects the experimental fact, that the carbynes
attached to graphene fragments can be bound with a variable number of
carbon atoms. Optical spectra of carbyne are related to the typical
$\pi$-bonds of carbon valence electrons, which are degenerate only in
the polyyne case. Electronic spectra have been measured before for
different polyyne molecules in neon matrices \cite{4} or in gas
phase.\cite{5}

The present article reports on the {\it ab initio} study of the
optical properties of cumulene and polyyne prototypes, within the
density-functional theory\cite{Hoh64,Koh65} (DFT) using the
independent-particle approximation\cite{Ehr59} (IPA) in order to
investigate if cumulenes and polyynes can be distinguished by means of
their optical spectra. The calculations have been carried out using
the ABINIT\cite{ABINIT} code. In order to support our findings, we
have compared them with results within the Random-Phase approximation
(RPA) including local-field (LF) effects performed using the YAMBO
code.\cite{YAMBO} We have derived our conclusions in two independent
ways: on one hand side we have compared the low- and medium-energy
spectral structure based on the DFT-IPA for our prototypes; on the
other hand side we have drawn conclusion based on general spectral
features which are obtained in both, the DFT-IPA and the RPA-LF
results.

Polyyne prototypes are simulated using short (seven or eight C) carbon
chains saturating the ends with one H atom at each side, while
cumulene prototypes are simulated with two H atoms terminating each
side; the latter case is also studied for different orientations of
the planes to which the terminating CH$_2$ belong, and the differences
in the spectra are reported. The terminating groups contribution has been
singled out by applying the so-called real-space cutoff
technique,\cite{Hog03,Cas03,Mon03} which was originally developed for
surfaces and which has been adapted here for molecules. Using this
technique, the influence of the H termination of the carbon chains on
the optical spectra has been determined.

This article is organized as follows: first, we introduce shortly
the methodological background on which our calculations are based as
well as the computational details (Sect.~\ref{Th}). Then we inspect
the groundstate properties of C$_7$H$_n$ and C$_8$H$_n$ chains
($n=2,4$) with various lengths and H arrangements, where we
investigate especially their bonding character and their magnetization
(Sect.\ref{GS}). The bonding character and the spin polarization are
reflected in their electronic properties (Sect.~\ref{Electronic}) on
which the final results, the absorption spectra, are based
(Sect.~\ref{Optical}). In the latter section, we discuss the spectral
structure obtained within DFT-IPA by performing a detailed peak
analysis as well as inspecting more general spectral
features. Besides, the influence of LF to the DFT-IPA is
discussed. Finally, we summarize and draw our conclusions.
%
%
%
\section{Theoretical background and Computational details}\label{Th}
In this section, the groundstate theory and the computational
details are summarized as well as the basics for the calculation of
optical spectra.
%
%
\subsection{Groundstate}
Our investigation starts from {\it ab initio} total energy calculations
using the ABINIT\cite{ABINIT,Gon05} code, which employs the
density-functional theory scheme\cite{Hoh64} (DFT) within the
local-density approximation\cite{Cep80,Per81} (LDA). The
eigenfunctions used to solve self consistently the Kohn-Sham
equations\cite{Koh65} are expanded in plane waves and, for the ionic
cores, we have chosen norm-conserving pseudopotentials in the
Troullier-Martins style.\cite{Tro91} ABINIT is designed for a
periodic-cell approach.  The molecule is placed in a supercell and
surrounded by vacuum to prevent the coupling to its periodic
images. Due to the 0-dimensionality of the system only the $\Gamma$
point is necessary to sample the reciprocal space. Therefore, the
convergence of the groundstate depends on the kinetic-energy cutoff
which determines the number of plane waves in the expansion and on the
size of the supercell.

For an accurate groundstate, convergence requires a kinetic-energy
cutoff of 20~Ha and a tetragonal supercell with lengths
$a=b=25$~$a_{\rm B}$ and $c=40$~$a_{\rm B}$. The carbon chain is
oriented along the $z$ direction ($c$ axis). This choice of the
convergence parameters yields total energy differences less than
$10^{-4}$~Ha. The carbynes have been relaxed till the remaining forces
are less than $5 \cdot 10^{-5}$~Ha/$a_{\rm B}$. For details, see
Ref.~\onlinecite{Mot08}.

YAMBO\cite{YAMBO,Mar2009} works directly on the ABINIT groundstate
results and no additional groundstate calculations are necessary.

%
%
\subsection{Optical properties}
The calculation of optical properties, in particular the
photo-absorption cross section $\sigma(\omega)$ as a function of the
energy $\omega$ is directly connected with the calculation of the
polarizability function $\alpha$, since\cite{Oni02}
\begin{eqnarray}
  \sigma(\omega)= \frac{\omega}{c_0} {\rm Im}[4\pi\bar{\alpha}(\omega)]
\end{eqnarray} 
holds. Herein, ${\rm Im}[\bar{\alpha}(\omega)]$ is the imaginary part of
the average (over the $x$, $y$, and $z$) polarizability function and
$c_0$ is the velocity of light. Thus it is sufficient to determine
${\rm Im}4\pi\bar{\alpha}(\omega)$.

In a first step, ${\rm Im}[4\pi\bar{\alpha}(\omega)]$ has been
calculated in the independent particle approach (IPA).\cite{Ehr59} The
probability ${P}_{v{\bf k},c{\bf k}}$ of the transitions between
valence ($v$) and empty ($c$) states with electronic eigenenergies
$E_{v{\bf k}}$ and $E_{c{\bf k}}$ at a given ${\bf k}$ point in the
reciprocal space can be calculated as the matrix elements of the
velocity operator. With these probabilities, ${\rm
  Im}[4\pi\alpha_{\nu}(\omega)]$ $(\nu=x, y, z)$ can be written
as\cite{Bas75}
\\ \parbox{7.5cm}{\begin{eqnarray*}
  {\rm Im} [ 4 \pi \alpha_{\nu}(\omega)]
  &=&
  \frac{8 \pi^2 e^2}{m^2 \omega^2}
  \sum_{\bf k}
  \sum_{v,c} \left|{P}^{\nu}_{v{\bf k},c{\bf k}}\right|^2 \\
  &&
  \times \delta(\Delta E 
  -\hbar \omega) \quad ,
 \end{eqnarray*}}\hfill
\parbox{8mm}{ \begin{eqnarray} \label{EqImalpha}\end{eqnarray}} \\
where $\Delta E = E_{c{\bf k}}-E_{v{\bf k}}$, and $m$ and $e$ are the
electronic mass and charge.  In this approximation, only the matrix
elements ${P}^{\nu}_{v{\bf k},c{\bf k}}$ and the electronic
eigenenergies are required. For molecules, the sum over {\bf k} can be
reduced to the use of the $\Gamma$ point only. Working within the
DFT-IPA, local-field, self-energy, and excitonic effects are
neglected.
 
In order to determine the influence of hydrogen termination of the
chains, we have employed the real-space cutoff technique (see, e.g.,
Hogan {\it et al.}\cite{Hog03}) to the DFT-IPA spectra. With this
technique the contribution of a selected region of the simulation cell
can be filtered out by including a boxcar function in the calculation
of the matrix elements. This function is equal to one in the desired
region and zero elsewhere. We have implemented the matrix elements
with and without real-space cutoff to the ABINIT\cite{ABINIT} code and
we have tested the implementation successfully.\cite{PREPRINT}

Beside the DFT-IPA, we have employed more elaborated methods in the
framework of the TDDFT in the linear response formalism.\cite{Mar2004}
In brief, this approach requires the solution of the Dyson equation
for the density response function. Here, also the standard
Random-Phase approximation (RPA) with the inclusion of local-field
(LF) effects is accessible. In order to obtain the polarizability
function for interacting particles (which is directly related to the
response function) one has to invert the Dyson equation. Usually, for
periodic cell systems this is done in reciprocal space
which can be quite cumbersome for supercell systems. Therefore, an alternative
method is provided by the Casida
approach,\cite{Cas1998a,Cas1998b,Jam1996,Cas2004,Vas1999} which moves
the task to the eigenvalue problem
\\ \parbox{6.0cm}{\begin{eqnarray*}
\left(\begin{array}{cc}
  A & B \\
  B & A \\ 
\end{array}\right)
= \omega
\left(\begin{array}{cr}
  1 & 0 \\
  0 & -1 \\ 
\end{array}\right)
\left(\begin{array}{c}
  x \\
  y \\ 
\end{array}\right)
\end{eqnarray*}}\hfill
\parbox{8mm}{ \begin{eqnarray} \label{Eq:TDDFT}\end{eqnarray}}\\
with $A=\Delta E+U_0+f_{\rm xc}$ and $B=U_0+f_{\rm xc}$, where $f_{\rm
  xc}$ is the exchange-correlation kernel and $U_0$ is the Coulomb
interaction term. For details see, e.g., Refs~\onlinecite{Cas2004} and
\onlinecite{Mar2004}. Eq.~(\ref{Eq:TDDFT}) is implemented in
YAMBO\cite{YAMBO,Mar2009} and allows the utilization of various
approximations. Setting $B=0$ yields the Tamm-Dancoff approximation
(TDA) whereas choosing $f_{\rm xc}=0$ the standard RPA-LF.

We have performed calculations using YAMBO\cite{YAMBO,Mar2009}
employing the RPA-LF using both, the Dyson equation solution as
well as the Casida approach.

%
%
%
\section{Groundstate properties}\label{GS}
\begin{figure}[t]
  \epsfig{figure=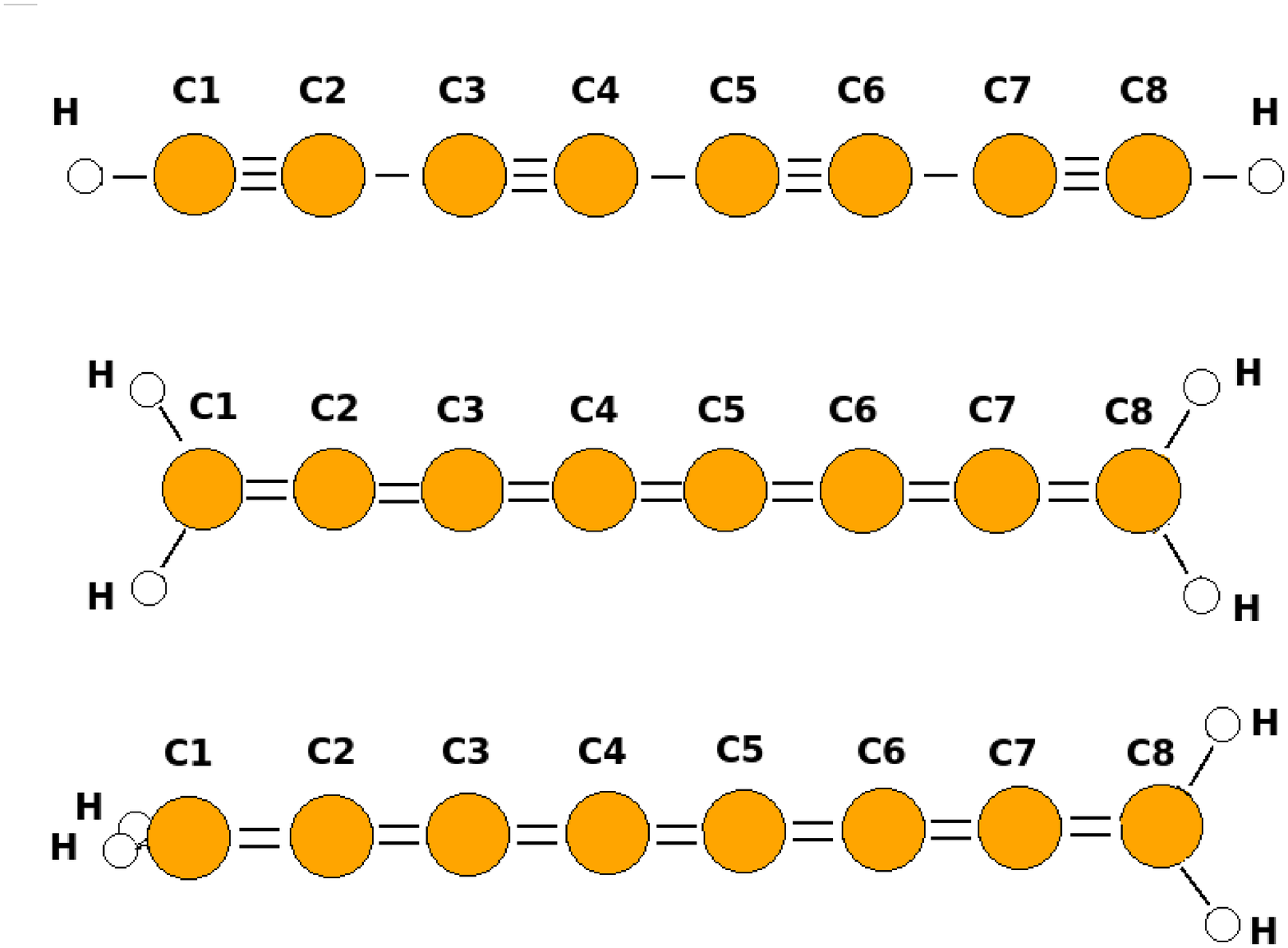,
    width=8.6cm,angle=0}
  \caption{(color online) Sketch of the carbynes C$_{8}$H$_{2}$ (upper
    panel), C$_{8}$H$_{4}$ in the $D_{2h}$ (middle panel) and in the
    $D_{2d}$ configuration (lower panel). Note the different bonding
    type of the molecules.
  }\label{FigSketch}
\end{figure}
In this study we consider the carbynes C$_{8}$H$_{2}$ and
C$_{8}$H$_{4}$ as representatives for even atomic polyynes and
cumulenes (C$_8$ chains). In the latter case, the terminating
hydrogens can have different orientations yielding the $D_{2h}$ and
$D_{2d}$ symmetry. The $D_{2h}$ has a planar configuration (all H and
the carbon chain itself are in one plane) while for the $D_{2h}$ one
pair of H atoms is placed in a plane perpendicular to the one
containing the other pair of H atoms, see Fig.~\ref{FigSketch}. The
same configurations are also studied for chains containing seven
carbon atoms (C$_7$ chains) as odd atomic prototypes. Besides, also
longer polyynes with an even number of carbon atoms are
inspected. Here we will discuss the groundstate properties in terms of
bond length and magnetization.

%
%
\subsection{Bond length}

\begin{figure}[t]
  \epsfig{figure=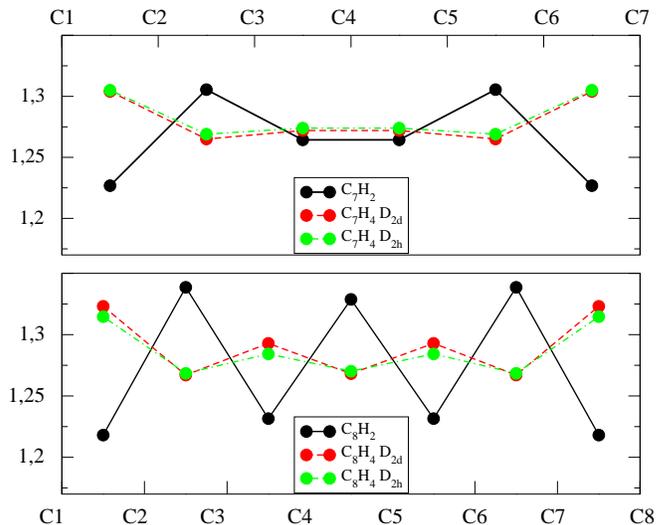,
    height=8.6cm,angle=-90}
  \caption{(color online) Equilibrium bond lengths of C$_{7}$ chains
    (upper panel) and C$_{8}$ ones (lower panel). 
  }\label{FigBonds}
\end{figure}

The relaxation of C$_{8}$H$_{4}$ results in nearly equal C-C distances
of 1.27--1.28~\AA\ due to the double-bond character of all internal
C-C bonds whereas for C$_{8}$H$_{2}$ the single and triple bonds
result in bond length of 1.33 and 1.23~\AA, respectively (see
Fig.~\ref{FigBonds}), as expected from the literature. The distances
C1-C2 and C7-C8 are slightly different due to the presence of the
terminating H atoms. As one can see in Fig.~\ref{FigBonds}, the
configuration ($D_{2h}$ or $D_{2d}$) of C$_{8}$H$_{4}$ does not
influence the bond length significantly. Since the bond length of
C$_{8}$H$_{4}$ are equal, no change is expected for longer
chains. Extending C$_{8}$H$_{2}$ to C$_{12}$H$_{2}$ and
C$_{16}$H$_{2}$, one observes that the length of the triple bonds does
not change whereas the average length of the single bonds is slightly
reduced to values around 1.31~\AA.

Inspecting C$_{7}$H$_{4}$ one obtains bond lengths very similar to the
case of C$_{8}$H$_{4}$ due to the double-bond character of the C-C
bindings. The situation is completely different for C$_{7}$H$_{2}$ due
to the odd number of C atoms. In C$_{8}$H$_{2}$ the C1-C2 and C7-C8
bonds at the end of the chain are both triple bonds and the
alternation of single and triple bonds continues for the whole chain
due to the even number of C atoms. For C$_{7}$H$_{2}$, the C-C bonds
at the end of the chain are triple bonds which is reflected in the
bond length of 1.23~\AA\ similar to the corresponding bond length of
C$_{8}$H$_{2}$. The neighboring bonds have a length of 1.31~\AA\ which
is between the length of the single and the double bonds of
C$_{8}$H$_{2}$. The chain ``tries'' to obtain alternating bond
length. In the middle of the chain, the alternation from both ends
gets in conflict and results in bonds with length of 1.26~\AA.

%
%
\subsection{Analysis of the magnetization}
\begin{figure}[t]
  \epsfig{figure=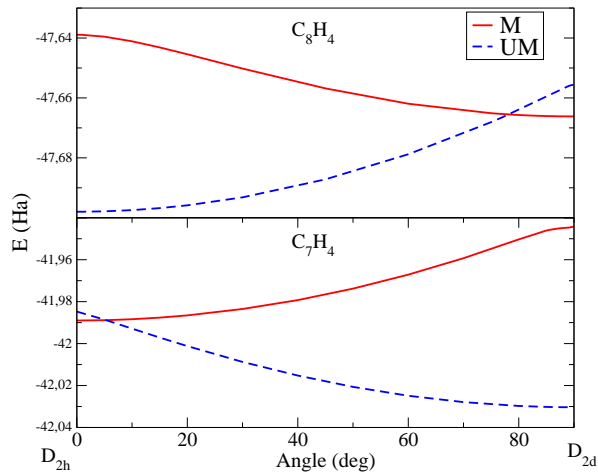,
    height=8.6cm,angle=-90}
  \caption{(color online) Groundstate energy of
    C$_{8}$H$_{4}$ (upper panel) and C$_{7}$H$_{4}$ (lower panel) as a
    function of the relative angle between the H-H terminations. The
    solid line marks the magnetic (M) and the dashed line the
    non-magnetic (UM) configuration. 
  }\label{FigGSspin}
\end{figure}

Analyzing the groundstate properties we obtained the expected result,
that C$_8$H$_4$ in the D$_{2d}$ configuration is magnetic, whereas
C$_8$H$_4$ D$_{2h}$ and C$_8$H$_2$ are not. Since the D$_{2d}$ differs
only by the angle between the planes in which the H atoms are placed,
we have investigated how the variation of angle between the planes
containing the CH$_2$ terminations affects the magnetization of the
cumulenes, varying the angle from 0$^\circ$ (D$_{2h}$) to 90$^\circ$
(D$_{2d}$). For this reason, we have performed total-energy
calculations forcing the Kohn-Sham orbitals to be occupied each with
two electrons (spin up and spin down) like in the non-magnetic case of
D$_{2h}$ and on the other hand side, we imposed the spin configuration
of the D$_{2d}$ cumulene (magnetic). With these two setups, we have
calculated the total energy as a function of the rotation angle. The
result for the C$_8$ chains is presented in Fig.~\ref{FigGSspin}
(upper panel). Note that in the case of the D$_{2h}$ configuration the
absence of spin polarization is energetically preferred by 59~mHa,
whereas for D$_{2d}$ the spin-polarized setup is more stable than the
spin-unpolarized one by 11~mHa. Turning the H terminations, the system
prefers to remain in the unpolarized configuration until an angle of
78$^\circ$.  After this angle the magnetic configuration becomes
energetically advantageous, as we found for C$_8$H$_4$ in the D$_{2d}$
symmetry. These findings are in agreement with the results of Ravagnan
et al.\cite{Rav09} performed with a different numerical implementation
of the DFT equations.

Interestingly, for the C$_7$ chains the situation is opposite to the
C$_8$ ones. Here, the C$_7$H$_2$ and C$_7$H$_4$ with D$_{2h}$ symmetry
are magnetic, whereas the C$_7$H$_4$ in the D$_{2d}$ configuration is
non-magnetic. Performing an analysis similar to the C$_8$H$_4$ case, one
finds that the D$_{2h}$ and the configurations until an angle of
5$^\circ$ are magnetic, while the configurations for larger angles and
the D$_{2d}$ are non-magnetic as shown in Fig.~\ref{FigGSspin} (lower
panel). Here, the energy difference between the magnetic and the
non-magnetic configuration for D$_{2h}$ is 4~mHa while it is 86~mHa
for D$_{2d}$.

%
%
%
\section{Electronic properties}\label{Electronic}

Optical transitions are excitations of electrons from a occupied state
to an empty one in case of absorption. This is the basic assumption for
DFT-IPA calculations. Furthermore, optical spectra can be interpreted
inspecting the electronic states with respect to their (relative)
energies and the charge density corresponding to these states. Thus,
it is useful to inspect the electronic properties of the carbynes
investigated here in detail. 

\begin{figure}[t]
  \epsfig{figure=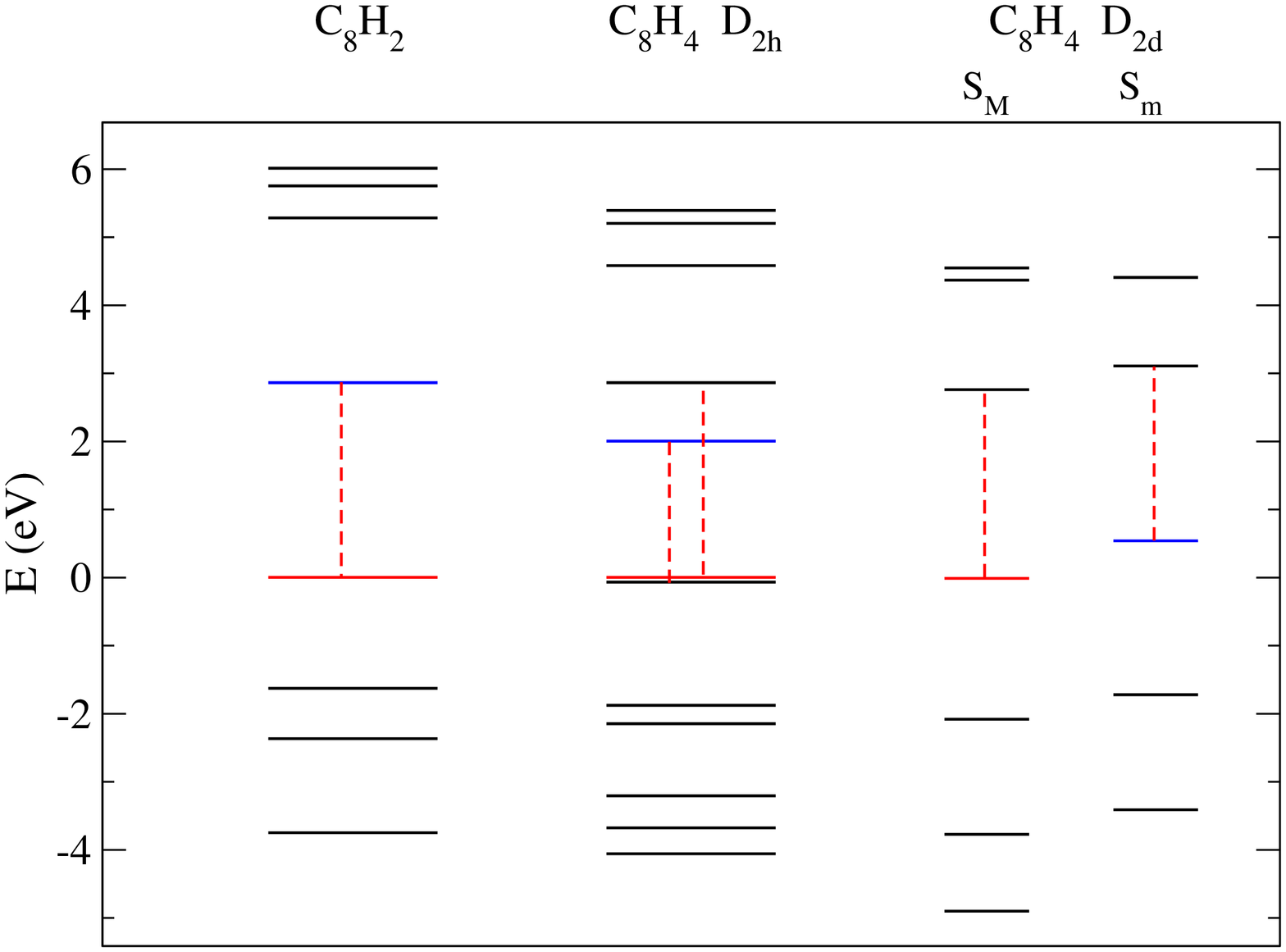,
    height=8.6cm,angle=90}
  \epsfig{figure=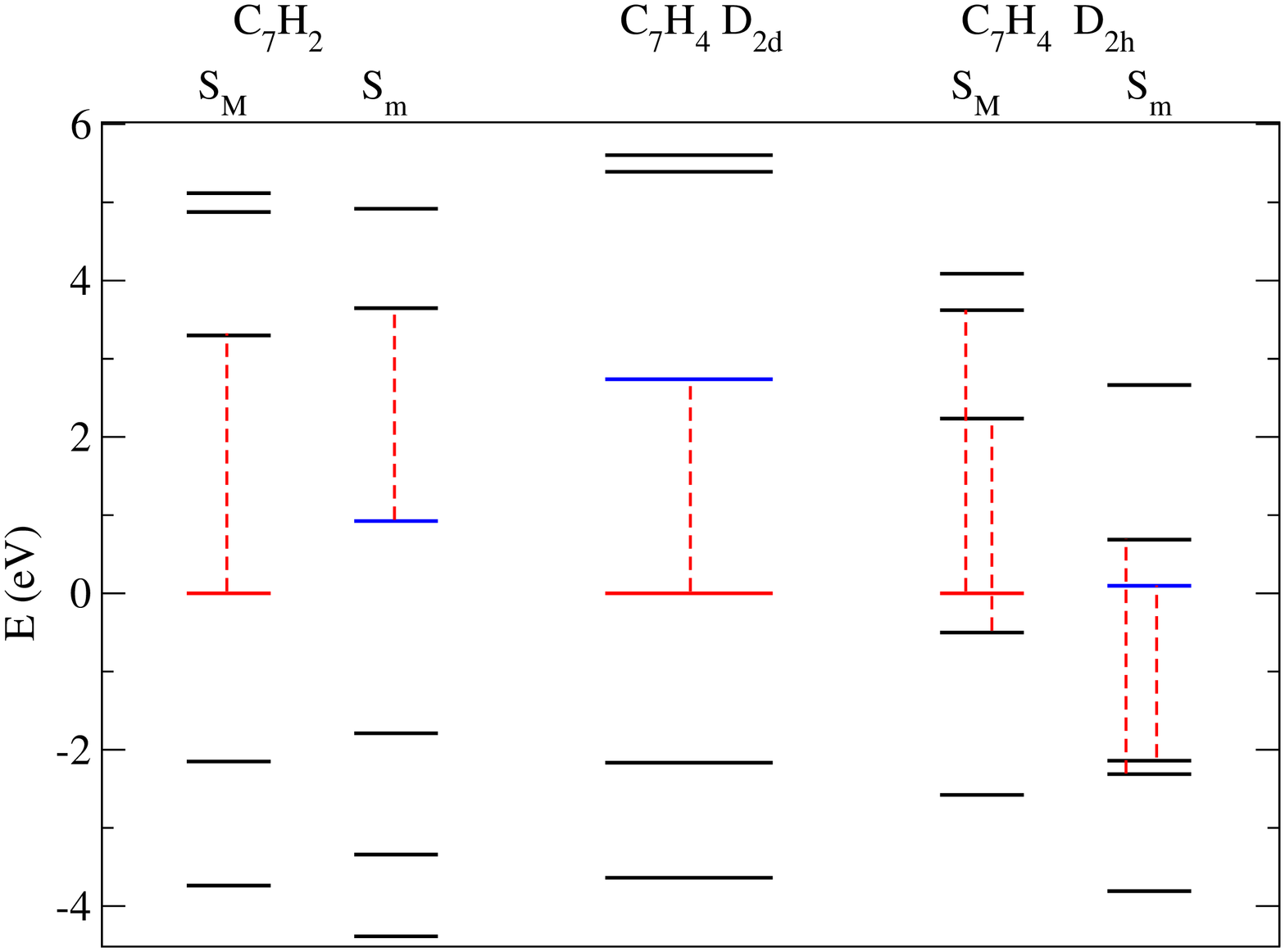,
    height=8.6cm,angle=-90}
  \caption{(color online) Schematic presentation of the relative
    positions of the DFT eigenenergies for the considered C$_8$ chains
    (upper panel) and the C$_7$ ones (lower panel). For magnetic
    chains, the eigenenergies for the major (M) and minor (m) spin is
    broken down. The energy scale is shifted to HOMO=0~eV (of the
    major spin for magnetic configurations). The main optical
    transitions within DFT-IPA are marked with dashed lines.
  }\label{FigBands}
\end{figure}

The calculated electronic eigenenergies for our carbyne prototypes are
schematically shown in Fig.~\ref{FigBands}. Since in the calculation
of DFT-IPA spectra only energy differences are considered, we have
shifted the energy range in such a way that the highest occupied
molecular orbital (HOMO) is at 0~eV. In case of spin polarization we
have chosen the HOMO of the major spin. The states close to the energy
gap between the HOMO and the lowest unoccupied molecular orbital
(LUMO) are energetically well distinguishable. Thus, one expects a
clear structure of non-overlapping peaks in the low-energy range of
the spectra.

\begin{figure}[t]
\begin{minipage}{8.6cm}
\fbox{\epsfig{figure=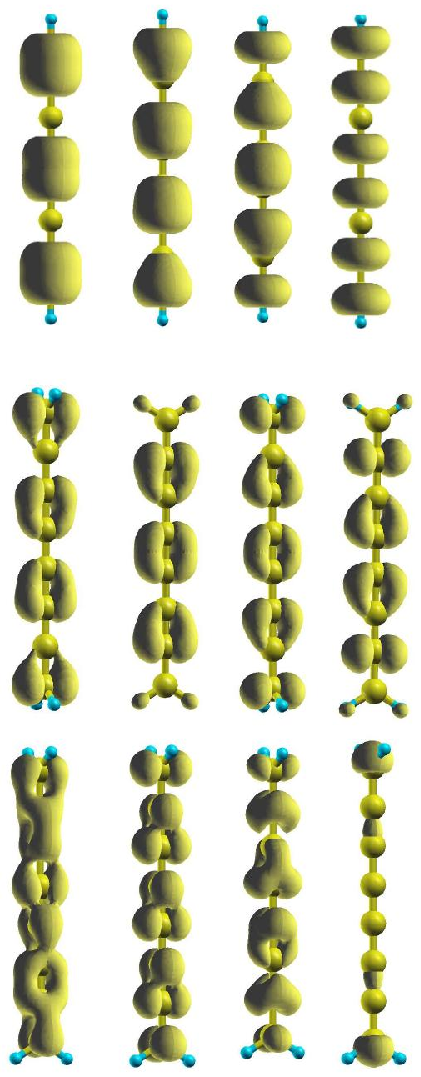, height=10cm, angle=0}}
\fbox{\epsfig{figure=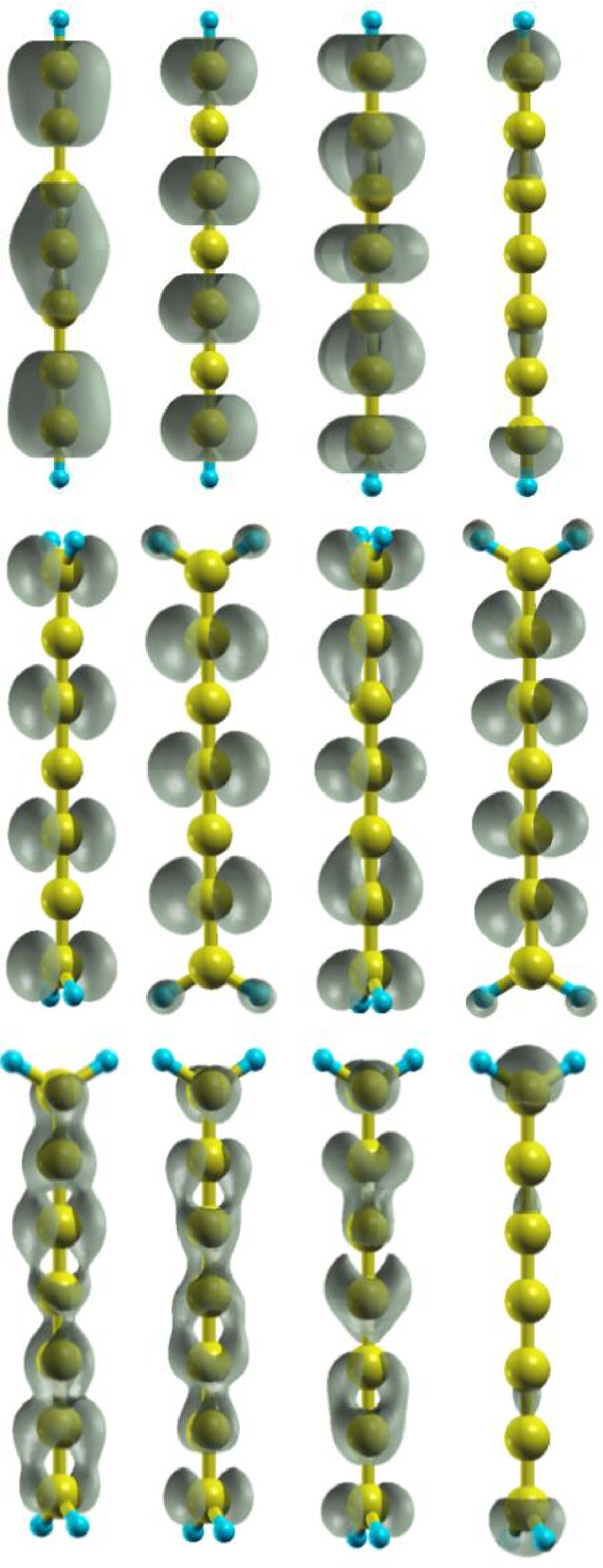, height=10cm, angle=0}}
\end{minipage}
  \caption{(color online) Charge density for C$_{8}$ chains (left
    panel) and C$_{7}$ ones (right panel); for each panel, from the
    fop to bottom, for the polyyne (upper panel), the D$_{2h}$
    cumulene (middle panel), and the D$_{2d}$ one the HOMO-1, HOMO,
    LUMO, and LUMO+1 are presented from left to right. 
  }\label{FigDensity}
\end{figure}

The electronic states of C$_8$H$_2$ close to the gap are double
degenerate due to the cylindrical symmetry of the chain. Because of
this, the coupled $\pi_{x}$ and $\pi_{y}$ single orbitals are at the
same energy. The triple and single bonds are clearly distinguishable
in charge density of the HOMO state (see Fig.~\ref{FigDensity}). The
linear $sp$ bonds are present at lower energies, since they are more
localized and feel more the attraction of the positive ions. LUMO
states, instead, are placed around the odd bonds. There is a strong
overlap between HOMO and LUMO around the carbons, in particular if we
consider C1 and C8. Because of this, one expects that a significant
contribution to the optical spectrum can be attributed to electronic
transitions in the spatial region of the last carbon atoms next to the
H terminations.

For C$_{8}$H$_{4}$ in the D$_{2h}$ configuration the cylindrical
symmetry is broken, so the electronic states are no more degenerate,
and there is an alternation of $x$ and $y$ polarized states. The HOMO
of C$_{8}$H$_{2}$ splits into C$_{8}$H$_{4}$ (D$_{2h}$) HOMO-1 and
LUMO, since the corresponding charge density is very similar. An
analogous situation occur for the LUMO of C$_{8}$H$_{2}$ (see
Fig.~\ref{FigDensity}), which splits into HOMO and LUMO+1. The overlap
between HOMO and LUMO is zero, since they are orthogonally polarized,
so we do not expect an absorption peak for a transition involving
these states. On the contrary, HOMO and LUMO+1 show a charge-density
distribution on the same plane. The same applies to HOMO-1 and
LUMO. For the latter states, we see in Fig.~\ref{FigDensity} an
important overlap around the termination, because these two states are
polarized orthogonally to the hydrogen plane. The contribution of this
region should be corrected by the real-space cutoff in order to
eliminate the H contribution to the spectra.

A mixed situation arises for C$_{8}$H$_{4}$ in the D$_{2d}$
symmetry. Here only the bands between HOMO-4 and LUMO are double
degenerate for spin up and the bands between HOMO-3 and LUMO+1 for
spin down. Majority and minority electron densities have a very
similar shape, therefore only one component is shown in the
Fig.~\ref{FigDensity}. The $\pi$-derived HOMO-1 can be understood as a
result of the merging of the HOMO-1 and HOMO of the planar
conformer.

Regarding the degeneracies and the splitting of eigenstates, the
C$_{8}$ and the C$_{7}$ chains are very similar. However, the
different bonding character of C$_{7}$H$_{2}$ is reflected in the
charge-density distribution of the HOMO and the HOMO-1. While for
C$_{8}$H$_{2}$ the charge density is placed around the triple bonds
for the HOMO, for the C$_{7}$H$_{2}$ it is localized at C
atoms. Especially in the HOMO-1, the difference due to the even/odd
number of C atoms can be seen: for C$_{7}$H$_{2}$ the corresponding
charge density covering the three C atoms localized in the center of
the molecule.

Looking forward to the calculation of the optical spectra and the use
of the real-space cutoff, the polyynes and the cumulenes should be
treated differently. Cumulenes are chains of the style ${\rm
  C}_{n-2}({\rm CH}_{2})_2$ and thus the termination of the chain
contains not only H but also C. Because of this, the cut should be
placed between the last two C atoms of the chain and cutting of a
complete CH$_2$ group. Inspecting the charge density for the states
close to the gap, only a little amount of charge is cut in the middle
and therefore, unphysical spectral features might be avoided. On the
contrary, for the polyynes the termination consists only of H
resulting in a cut between the C and H atom. As shown in
Fig.~\ref{FigDensity}, for states close to the HOMO-LUMO gap, nearly
no charge density would be cut.

In general, the charge-density distribution is very regular leading
the conclusion that a similar distribution will be found even for
longer chains. The only chain where charge-density differences might
be found is C$_{7}$H$_{2}$ where the regimes showing alternating bonds
and double bonds have different extents for longer chains.

%
%
%
\section{Optical properties}\label{Optical}
In this section we present our results for the optical spectra
of our polyyne and cumulene prototypes. In a first step, we analyze
the spectral structure of our DFT-IPA spectra where we compare the
peak positions in the low- and medium-energy range
(Sect.~\ref{AbsDFTIPA}). Here, we have investigated also the
contribution of the terminating H atoms to the spectra using the
real-space cutoff technique (Sect.~\ref{AbsCut}).

Since the DFT-IPA yields fast but not highly accurate results, an
alternative scheme for distinguishing the considered six prototypical
chains has been developed based on general features observed in
the optical spectra (Sect.~\ref{AbsGENERAL}). In order to confirm the
basic ingredients for this scheme, we have performed RPA-LF
calculations for the non-magnetic systems C$_8$H$_2$, C$_8$H$_4$
(D$_{2h}$), and C$_7$H$_4$ (D$_{2d}$) (Sect.~\ref{AbsRPALF}).

%
%
\subsection{Spectra within DFT-IPA}\label{AbsDFTIPA}
For the calculation of the imaginary part of the dielectric function
it was possible to reduce the kinetic-energy cutoff to 18~Ha without
loosing significant accuracy (for detail with respect to this
parameter, see Ref.~\onlinecite{PREPRINT}).  This leads to a faster
but equally accurate calculation.  Another important parameter is the
number of empty states to be considered (see Eq.~\ref{EqImalpha}). For
molecules the continuum states should be excluded, since an excitation
of the molecule to these states corresponds to ionization. Continuum
states are extended in space and thus, the eigenenergies corresponding
to these states vary with the size of the supercell. An inspection of
the energy of many states as a function of the size of the supercell
yields the result that the energy of the states up to LUMO+7 is not
influenced significantly by the size of the simulation box while the
others are. Thus, we have considered in our calculation 8 empty
states. Beside the parameters mentioned here, all other parameters are
kept the same as for the groundstate calculations.

\begin{figure}[t]
  \epsfig{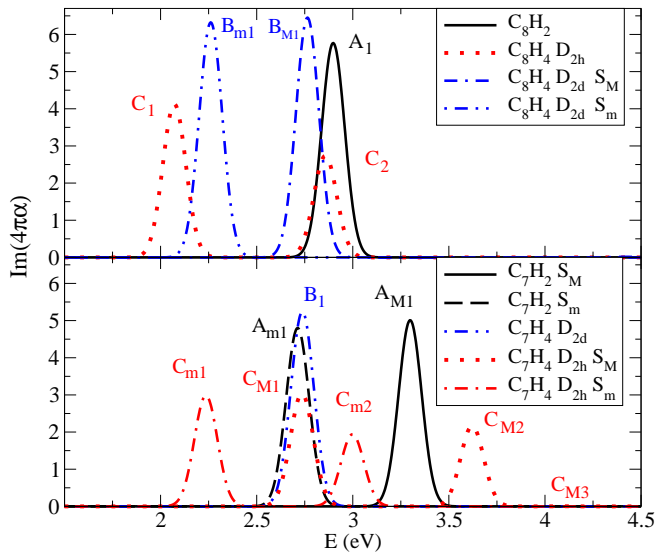}
  \caption{(color online) Comparison of the imaginary part of the
    polarizability function as a function of the energy for the C$_8$
    chains (upper panel) and the C$_7$ chains (lower panel). S$_{\rm
      M}$ correspond to major spin, S$_{\rm m}$ to minor spin
    spectra. The peaks are labeled with A for polyynes, B for D$_{2d}$
    cumulenes, and C for D$_{2h}$ cumulenes.  Here only the low-energy
    range is shown.
  }\label{FigAbsLow}
\end{figure}

\begin{figure}[t]
  \epsfig{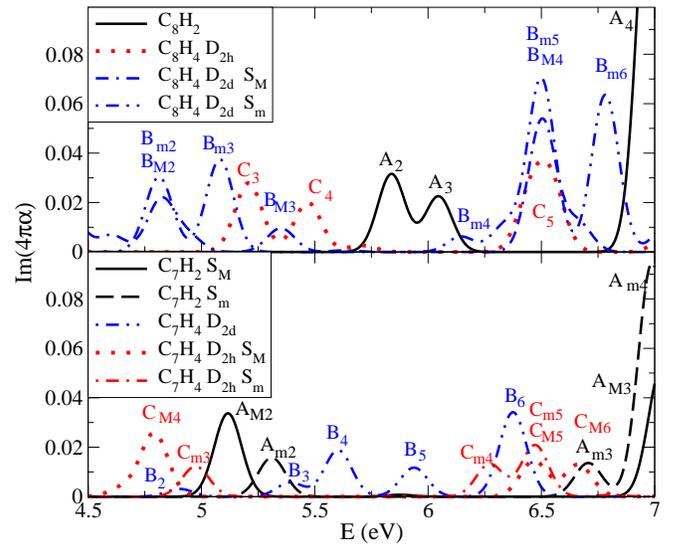}
  \caption{(color online) Same as in Fig~\ref{FigAbsLow} for the
    medium-high energy range.
  }\label{FigAbsMed}
\end{figure}

The calculated spectra of the imaginary part of the polarizability
function as a function of the energy for C$_8$H$_2$, C$_8$H$_4$
(D$_{2h}$ and D$_{2d}$), C$_7$H$_2$, and C$_7$H$_4$ (D$_{2h}$ and
D$_{2d}$) are presented in Fig.~\ref{FigAbsLow} (low-energy range) and
Fig.~\ref{FigAbsMed} (medium-energy range), where also the resolution
in major and minor spin contributions is given for magnetic compounds.

Inspecting the low-energy range of the spectra of C$_8$H$_2$
(Fig.~\ref{FigAbsLow}), one notice immediately that the three
considered prototypes are clearly distinguishable by their spectral
structure: C$_{8}$H$_{2}$ has only one strong peak (marked with A$_1$
in the figure) at 2.90~eV which results from transitions between the
HOMO and LUMO states. This transition involves electrons in the
$\pi_{x}$ and $\pi_{y}$ molecular orbitals (see
Fig.~\ref{FigDensity}), which are degenerate due to cylindrical
symmetry. Thus, the peak is polarized in $z$ direction. For the
non-magnetic C$_8$H$_4$ D$_{2h}$ cumulene the main peak is split into
two peaks, whose energies are 2.07~eV and 2.85~eV. Since the
cylindrical symmetry is broken, both the HOMO and LUMO states of
C$_{8}$H$_{2}$ are split in two states for C$_8$H$_4$ (see
Fig.~\ref{FigBands}). Here the HOMO$\to$LUMO transition is forbidden
because the respective wave functions lie in the $x$-$z$ and $y$-$z$
planes. The peaks are due to the transitions HOMO-1$\to$LUMO (2.07~eV,
C$_{1}$) and HOMO$\to$LUMO+1 (2.85~eV, C$_{2}$). The C$_8$H$_4$
D$_{2d}$ shows also a double-peak structure, however, one peak is due
to a major-spin (2.76~eV, B$_{M1}$) and the other due to a minor-spin
(2.26~eV, B$_{m1}$) transition. Both transitions are HOMO-LUMO
ones. From the low-energy range of the DFT-IPA spectra one can
conclude that the C$_8$ chains can be distinguished in optical
spectra, since the polyyne prototype shows only one main peak whereas
the cumulenes show a double-peak structure. The two cumulenes can be
distinguished in symmetry due to the fact that one is magnetic and the
other not.

Since the C$_{8}$H$_{2}$ polyyne might have not exactly the
cylindrical symmetry in the experiment, we have also calculated the
groundstate and the spectrum of a non-linear configuration of
C$_{8}$H$_{2}$ where we altered the carbon positions in order to
create a zig-zag configuration shifting the carbons of the molecule
(which is oriented along the $z$ axis) in the $x$ or $y$ direction by a
displacement of about the 5~\% of bond lengths. In this way we have
broken the cylindrical symmetry. Because of this, the main peak A$_1$
is split in two close peaks at 2.85~eV and 2.80~eV. For peaks at
higher energy we have found that the higher is the energy of the peak,
the higher is its energy splitting. Nevertheless, this split is much
less than the ones of the cumulenes (C$_{1}$, C$_{2}$ and B$_{M1}$,
B$_{m1}$, respectively) and thus, the conclusion from above remains
valid also in this case.

Inspecting the spectra of the C$_7$ chains, one observes a clear
spectral structure as in the C$_8$ case. C$_{7}$H$_{2}$ is magnetic
and the HOMO-LUMO transition splits into two peaks at 3.30~eV
(A$_{M1}$, major spin) and at 2.71~eV (A$_{m1}$, minor
spin). Similarly, the non-magnetic C$_{7}$H$_{4}$ (D$_{2d}$) shows
only one peak at 2.74~eV (B$_1$) due to the HOMO-LUMO transition. This
peak appeared split for the magnetic C$_{8}$H$_{4}$ D$_{2d}$
cumulene. The magnetic C$_{7}$H$_{4}$ D$_{2h}$ shows four transitions:
for the major spin we have the HOMO-1$\to$LUMO (2.74~eV, C$_{M1}$) and
the HOMO$\to$LUMO+1 (3.62~eV, C$_{M2}$) one and for the minor spin a
HOMO$\to$LUMO (2.23~eV, C$_{m1}$) and a HOMO-1$\to$LUMO+1 (2.99~eV,
C$_{m1}$) one. As for the C$_8$ case, comparing the C$_7$ chains
one can conclude that the three prototypes can be distinguished
since we have four peaks for C$_{7}$H$_{4}$ D$_{2h}$, two peaks for
C$_{7}$H$_{2}$, and one peak for C$_{7}$H$_{4}$ (D$_{2d}$).

Comparing the DFT-IPA spectra for the C$_7$ and the C$_8$ chains, from
the spectral structure all types of chains could be identified in
experiments if their magnetization is known. The four-peak feature of
C$_{7}$H$_{4}$ D$_{2h}$ is unique. We have two magnetic double-peak
features (C$_{8}$H$_{4}$ D$_{2d}$ and C$_{7}$H$_{2}$) which are
distinguishable due to the transition energy. We have one unique
non-magnetic double peak for C$_{8}$H$_{4}$ D$_{2h}$. The only chains
which cannot be distinguished unambiguously from the low-energy range
of the DFT-IPA spectra are C$_{8}$H$_{2}$ and C$_{7}$H$_{4}$ D$_{2d}$
which are both non-magnetic and show only a single peak which differs
in energy only about 0.1~eV. Note, that in the low-energy range all peaks shown in Fig.~\ref{FigAbsLow}
are $z$ polarized.

Unfortunately, DFT-IPA spectra usually can give only a rough idea of
what might be seen in the experiment. Comparing spectra performed with
TDDFT methods or those which are based on solutions of the
Bethe-Salpeter equation (for an overview of methods and results see,
e.g., Ref.~\onlinecite{Oni02}), one observes not only a shift of the
peaks but also a shift of the oscillator strength to higher energies
which is not uniform. Nevertheless, the low-energy transitions will
contribute also to the corresponding peaks at TDDFT level, though,
possibly at less extent. Regarding the shift of peak positions and
oscillator strength to higher energies, the medium and high-energy
range can be enhanced. Therefore, we inspect the medium-energy range
of our DFT-IPA spectra, too.

For the medium energy range up to 7~eV (see Fig.~\ref{FigAbsMed}) a
clear spectral structure is absent as well as for the high-energy
range (not shown here). In this range, the DFT-IPA spectra shows only
a low intensity, which is of about $0.5\%-2.5\%$ of the intensity in
the low-energy range. The spectral features involve transitions from
states down to HOMO-5 to states up to LUMO+5. Inspecting the
transitions, no general trends have been found. Considering the
polarization of the transitions, most of the peaks are $z$ polarized. The remaining peaks have an $x$ and/or $y$
component only. Considering the magnetic compounds, one sees that some
of the peaks for major and minor spin coincide in energy.  For
C$_{8}$H$_{4}$ D$_{2d}$ we have two of these peaks, B$_{M2/m2}$ and
B$_{m4}$/B$_{M4}$.  For C$_{7}$H$_{4}$ D$_{2h}$ it is only one, namely
C$_{M5/m5}$ and for C$_{7}$H$_{2}$ it is A$_{M3/m4}$.

%
%
\subsection{Spectra using the real-space cutoff}\label{AbsCut}
The DFT-IPA technique gives us the possibility to determine the
contribution of the terminating H atoms to the spectra by employing
the real-space cutoff introduced by various groups, see e.g.,
Refs.~\onlinecite{Hog03,Cas03,Mon03}.  As discussed in
Sect.~\ref{Electronic}, the best choice of eliminating the H
termination would be to cut the H atoms only for polyynes and the
terminating CH$_2$ group for cumulenes. However, the resulting chains
would not be comparable in intensity due to the different number of
remaining C atoms. Thus, we have decided to eliminate the terminating
H atoms together with the neighboring C atom (resulting in the CH$_2$
group for cumulenes and a CH group for polyynes) by
applying the real-space cutoff for the polyynes and cumulenes in the
same range.

Inspecting the low-energy range, we have found a lowering of around
25\% of the main peaks for the C$_8$ chains applying the real-space
cutoff. This can be explained by the fact that we have cut also one C
atom at each end and thus, the contribution of the H atoms is very
small. We can also conclude, that the intensity of the peaks is
determined by the number of C atoms in the chain which is confirmed by
comparing the intensities of the C$_8$ and the C$_7$ chains in
Fig.~\ref{FigAbsLow}. Thus, one expects for longer chains not only
spectra containing more peaks (since there are more non-continuum
states to consider) but also with a larger intensity.

%
%
\subsection{Conclusions from general considerations}\label{AbsGENERAL}
Often DFT-IPA spectra are not able to predict a quantitative correct
peak structure. Thus, we have analyzed our spectra also from a more
general point of view in order to find an alternative way to draw
conclusion. At higher energies mainly $x$ and $y$ polarized peaks are
observed, whereas only few $z$ polarized feature appear. Looking at
the $z$ polarized peaks only, we have found a gap of at least 2~eV
between the last $z$ polarized peaks and the following high-energy $z$
polarized feature. This inset of the gap is placed at around 8-10~eV
and depends on the kind of the chain. Similarly, in the low- and
medium-energy range up to about 9~eV only few peaks with $x$ and/or
$y$ component are found.  This brings us to the conclusion that the
spectra can be divided in energy range: in the low and medium energy
range the $z$ component plays a major role whereas in the high-energy
range the spectra is determined by $x$ and $y$ polarized peaks. This
observation is useful for further considerations, since it allows the
counting of $z$ polarized peaks in the low-energy range till the gap
inset.

Inspecting other general features of the chains like symmetry and
magnetization, we are able to draw conclusion without detailed peak
analysis.

In a first step, the six chains can be divided into two groups, the
magnetic chains C$_8$H$_4$ (D$_{2d}$), C$_7$H$_2$, and C$_7$H$_4$
(D$_{2h}$) and the non-magnetic chains C$_8$H$_2$, C$_8$H$_4$
(D$_{2h}$), and C$_7$H$_4$ (D$_{2d}$), which can be discriminated by
the presence of a magnetic moment.

The second step is guided by symmetry considerations: in each group
there is only one chain which has no $xy$ symmetry: the magnetic
C$_8$H$_4$ (D$_{2h}$) and the non-magnetic C$_7$H$_4$ (D$_{2h}$) where
the $x$ and $y$ components of the spectra differ. For these chains one
expects spectral changes (in the high-energy regime where $x$ and $y$
polarized features appear) for $x$ and $y$ polarized light.

For the last step we exploit the fact that the spectra can be divided
into two zones: the range up to 8-10~eV where the spectra shows mainly
$z$ polarization and the range beyond 8-10~eV, where the spectra has
mainly $x$ and $y$ polarization. This range is separated by a gap of
at least 2~eV from the next small $z$ polarized peak. Since for the
polyyne chains some states close to the HOMO-LUMO gap are degenerate,
one expects less peaks for C$_7$H$_2$ and C$_8$H$_2$ than for
C$_8$H$_4$ (D$_{2d}$) and C$_7$H$_4$ (D$_{2d}$) in the magnetic and
non-magnetic case, respectively. Since the spectra in this range show
mainly $z$ polarized peaks, we are able to count these peaks till the
inset of the gap: in the energy range up to about 10~eV we have three
peaks for C$_8$H$_2$, 6 peaks for C$_8$H$_4$ (D$_{2h}$), 5 peaks for
C$_8$H$_4$ (D$_{2d}$ S$_{M}$), and 7 peaks for C$_8$H$_4$ (D$_{2d}$
S$_{m}$). Similarly, there are 2 peaks for C$_7$H$_2$ (S$_{M}$),
3 peaks for C$_7$H$_2$ (S$_{m}$), 4 peaks for C$_7$H$_4$
(D$_{2d}$), 4 peaks for C$_7$H$_4$ (D$_{2h}$ S$_{M}$), and 6
peaks for C$_7$H$_4$ (D$_{2h}$ S$_{m}$). For the comparison of
C$_7$H$_2$ with C$_8$H$_4$ (D$_{2d}$) (both magnetic) this means that
we have 2/3 (S$_{M}$/S$_{m}$) against 5/7 (S$_{M}$/S$_{m}$) or in
total 6 against 8, while for C$_8$H$_2$ and C$_7$H$_4$ (D$_{2d}$)
there are 3 against 4 peaks.

With these simple considerations we have shown, that the 6 chains can
be distinguished by optical spectra without the necessity of a
detailed peak analysis. 

\section{Spectra using RPA-LF}\label{AbsRPALF}

In order to inspect spectral changes using more advanced methods, we
have performed calculations using the RPA-LF. The main scope here is
to confirm the validity of the general conclusions, in particular, the
existence of the gap which is used to design a criterium for the peak
counting and the number of peaks in the low-and medium energy
range. For this reason, it was sufficient to do calculations not at
full convergence, which is difficult to achieve. For this test we
have performed calculations for the non-magnetic chains C$_8$H$_2$,
C$_8$H$_4$ (D$_{2h}$), and C$_7$H$_4$ (D$_{2d}$) only.

The results show mainly the expected shift of peak positions and
oscillator strength to higher energies. In detail, we have found a
strong shift to higher energies of the peaks shown in
Fig.~\ref{FigAbsLow}, which now appear at energies around 5-6~eV, and
an enhancement of the peaks with $z$ polarization shown in
Fig.~\ref{FigAbsMed}. The shift is not uniform for the low-energy
peaks. For example we have found for C$_8$H$_4$ only one strong peak
followed by three small peaks whereas at DFT-IPA there are two strong
peaks. Instead for C$_7$H$_4$ a double-peak structure with strong
intensity appear followed by a less intense double-peak structure. It
is possible that the first strong-intensity double feature merge to
one peak at higher convergence. For C$_8$H$_2$ a strong peak remained
followed by two small ones. Nevertheless, the three non-magnetic
chains still show a different spectral structure and are therefore
distinguishable by means of optical measurements. 

Regarding the general conclusions of Sect.~\ref{AbsGENERAL}, we have
found that for the $z$ polarized peaks the oscillator strength are
shifted from the high-energy peaks (above the gap) to the low-energy
ones, while the $x$ and $y$ component remained nearly unchanged in
intensity. Since the high-energy $z$ polarized peaks have a very small
intensity, a considerable spectral structure has been found only up to
8/9~eV followed by a gap of 2~eV. Thus, a counting of the peaks is
also possible here. In particular, we have found 3 peaks for
C$_8$H$_2$ compared to 3 peaks DFT-IPA, four peaks for C$_8$H$_4$
(D$_{2h}$) compared to 6 peaks DFT-IPA, and 5 peaks (2 strong
double peaks followed by a tiny one) for C$_7$H$_4$ (D$_{2d}$)
compared to 4 peaks DFT-IPA. Also here, C$_8$H$_2$ and C$_7$H$_4$
(D$_{2d}$) are distinguishable by the number of $z$ polarized peaks.

In summary, employing the RPA-LF we find similar spectral features as
in the DFT-IPA which have been used for our general conclusions.

%
%
%
\section{Summary and Conclusion}\label{Summary}
We have performed an {\it ab initio} investigation of carbyne
prototypes inspecting their groundstate and optical properties in
detail using DFT based methods. We have considered seven- and
eight-atomic chains saturated with one H (polyyne prototype) and two H
(cumulene prototype) at each end, respectively. Polyynes and cumulenes
have a different bonding character. While the cumulenes show always
double bonds, the polyynes have an alternating triple-single bond
character. In case of odd-atomic chains, we have observed that this
alternation results in the accidental building bonds similar to double
bonds in the middle of the chain. Regarding the cumulenes, there are two possible symmetry configuration, namely
D$_{2h}$ and a D$_{2d}$, corresponding to placing the
saturating H atoms all in one plane or placing them in planes
perpendicular to each other. Depending on the symmetry, the C$_8$H$_4$
D$_{2d}$ is magnetic and C$_8$H$_4$ D$_{2h}$ is not. For C$_7$H$_4$
the situation is reversed: here the D$_{2h}$ chain is magnetic whereas
C$_7$H$_4$ is non-magnetic. Rotating the H atoms, we have observed
that the angle for which the cumulene becomes magnetic is different
for even- and odd-atomic chains. Considering the optical spectra, we
have found that the H terminations give only tiny contributions to the
spectra. From the analysis of our DFT-IPA spectra we have seen that
the six carbynes considered here are distinguishable by their optical
spectra. However, due to well-known problems of the DFT-IPA a
quantitative conclusion cannot be drawn unambiguously. Nevertheless,
we have been able to draw a scheme to distinguish the chains
unambiguously based on the general features of the calculated spectra
(within DFT-IPA and RPA-LF), the spin polarization of the systems, and
their symmetry. With this scheme, the carbynes C$_{8}$H$_{2}$,
C$_{8}$H$_{4}$ D$_{2h}$, C$_{8}$H$_{4}$ D$_{2d}$, C$_{7}$H$_{2}$,
C$_{7}$H$_{4}$ D$_{2h}$, and C$_{7}$H$_{4}$ D$_{2d}$ can be
distinguished. It might be possible to extend this scheme also for
longer chains.

To apply our scheme, one needs the knowledge of the magnetization of
the system and polarization resolved optical spectra are
necessary. However, more accurate calculations might give quantitative
spectra from which the chains can be distinguished by absorption
spectra only without inspection of the magnetization of the chains
and/or polarization resolution of the spectra. These calculations
might be accelerated due to the fact, that the low-energy range of the
spectra is mainly $z$ polarized and thus a detailed inspection of the
$x$ and $y$ component is not necessary. We hope that this study here
will encourage further theoretical and experimental investigations of
carbynes.

%
%
%
\section*{Acknowledgment}
The authors thank Giovanni Onida, Silvana Botti, and Davide Sangalli
for fruitful discussions. This work was funded by the EU's 6th
Framework Programme through the NANOQUANTA Network of Excellence
(NMP-4-CT-2004-500198) and the EU's 7th Framework Programme through
the e-I3 ETSF initiative (Grant agreement No. 211956).
%
%
%
\bibliography{Theory,Opt,carlo,MyAndOthers.bib}

\begin{thebibliography}{37}
\expandafter\ifx\csname natexlab\endcsname\relax\def\natexlab#1{#1}\fi
\expandafter\ifx\csname bibnamefont\endcsname\relax
  \def\bibnamefont#1{#1}\fi
\expandafter\ifx\csname bibfnamefont\endcsname\relax
  \def\bibfnamefont#1{#1}\fi
\expandafter\ifx\csname citenamefont\endcsname\relax
  \def\citenamefont#1{#1}\fi
\expandafter\ifx\csname url\endcsname\relax
  \def\url#1{\texttt{#1}}\fi
\expandafter\ifx\csname urlprefix\endcsname\relax\def\urlprefix{URL }\fi
\providecommand{\bibinfo}[2]{#2}
\providecommand{\eprint}[2][]{\url{#2}}

\bibitem[{\citenamefont{Kawai et~al.}(2005)\citenamefont{Kawai, Okada,
  Miyamoto, and Oshiyama}}]{1}
\bibinfo{author}{\bibfnamefont{T.}~\bibnamefont{Kawai}},
  \bibinfo{author}{\bibfnamefont{S.}~\bibnamefont{Okada}},
  \bibinfo{author}{\bibfnamefont{Y.}~\bibnamefont{Miyamoto}}, \bibnamefont{and}
  \bibinfo{author}{\bibfnamefont{A.}~\bibnamefont{Oshiyama}},
  \bibinfo{journal}{Phys. Rev. B} \textbf{\bibinfo{volume}{72}},
  \bibinfo{pages}{035428} (\bibinfo{year}{2005}).

\bibitem[{\citenamefont{Darancet et~al.}(2009)\citenamefont{Darancet, Olevano,
  and Mayou}}]{2}
\bibinfo{author}{\bibfnamefont{P.}~\bibnamefont{Darancet}},
  \bibinfo{author}{\bibfnamefont{V.}~\bibnamefont{Olevano}}, \bibnamefont{and}
  \bibinfo{author}{\bibfnamefont{D.}~\bibnamefont{Mayou}},
  \bibinfo{journal}{Phys. Rev. Lett.} \textbf{\bibinfo{volume}{102}},
  \bibinfo{eid}{136803} (\bibinfo{year}{2009}).

\bibitem[{\citenamefont{Kong et~al.}(2000)\citenamefont{Kong, Franklin, Zhou,
  Chapline, Peng, Cho, and Dai}}]{3}
\bibinfo{author}{\bibfnamefont{J.}~\bibnamefont{Kong}},
  \bibinfo{author}{\bibfnamefont{N.~R.} \bibnamefont{Franklin}},
  \bibinfo{author}{\bibfnamefont{C.}~\bibnamefont{Zhou}},
  \bibinfo{author}{\bibfnamefont{M.~G.} \bibnamefont{Chapline}},
  \bibinfo{author}{\bibfnamefont{S.}~\bibnamefont{Peng}},
  \bibinfo{author}{\bibfnamefont{K.}~\bibnamefont{Cho}}, \bibnamefont{and}
  \bibinfo{author}{\bibfnamefont{H.}~\bibnamefont{Dai}},
  \bibinfo{journal}{Science} \textbf{\bibinfo{volume}{287}},
  \bibinfo{pages}{622} (\bibinfo{year}{2000}).

\bibitem[{\citenamefont{Lucotti et~al.}(2006)\citenamefont{Lucotti, Tommasini,
  Zoppo, Castiglioni, Zerbi, Cataldo, Casari, Bassi, Russo, Bogana
  et~al.}}]{Lu05}
\bibinfo{author}{\bibfnamefont{A.}~\bibnamefont{Lucotti}},
  \bibinfo{author}{\bibfnamefont{M.}~\bibnamefont{Tommasini}},
  \bibinfo{author}{\bibfnamefont{M.~D.} \bibnamefont{Zoppo}},
  \bibinfo{author}{\bibfnamefont{C.}~\bibnamefont{Castiglioni}},
  \bibinfo{author}{\bibfnamefont{G.}~\bibnamefont{Zerbi}},
  \bibinfo{author}{\bibfnamefont{F.}~\bibnamefont{Cataldo}},
  \bibinfo{author}{\bibfnamefont{C.}~\bibnamefont{Casari}},
  \bibinfo{author}{\bibfnamefont{A.~L.} \bibnamefont{Bassi}},
  \bibinfo{author}{\bibfnamefont{V.}~\bibnamefont{Russo}},
  \bibinfo{author}{\bibfnamefont{M.}~\bibnamefont{Bogana}},
  \bibnamefont{et~al.}, \bibinfo{journal}{Chem. Phys. Lett.}
  \textbf{\bibinfo{volume}{417}}, \bibinfo{pages}{78} (\bibinfo{year}{2006}).

\bibitem[{\citenamefont{Gu et~al.}(2008)\citenamefont{Gu, Kaiser, and
  Mebel}}]{Gu2008}
\bibinfo{author}{\bibfnamefont{X.}~\bibnamefont{Gu}},
  \bibinfo{author}{\bibfnamefont{R.}~\bibnamefont{Kaiser}}, \bibnamefont{and}
  \bibinfo{author}{\bibfnamefont{A.}~\bibnamefont{Mebel}},
  \bibinfo{journal}{Chem. Phys. Chem.} \textbf{\bibinfo{volume}{9}},
  \bibinfo{pages}{350} (\bibinfo{year}{2008}).

\bibitem[{\citenamefont{Ravagnan et~al.}(2009)\citenamefont{Ravagnan, Manini,
  Cinquanta, Onida, Sangalli, Motta, Devetta, Bordoni, Piseri, and
  Milani}}]{Rav09}
\bibinfo{author}{\bibfnamefont{L.}~\bibnamefont{Ravagnan}},
  \bibinfo{author}{\bibfnamefont{N.}~\bibnamefont{Manini}},
  \bibinfo{author}{\bibfnamefont{E.}~\bibnamefont{Cinquanta}},
  \bibinfo{author}{\bibfnamefont{G.}~\bibnamefont{Onida}},
  \bibinfo{author}{\bibfnamefont{D.}~\bibnamefont{Sangalli}},
  \bibinfo{author}{\bibfnamefont{C.}~\bibnamefont{Motta}},
  \bibinfo{author}{\bibfnamefont{M.}~\bibnamefont{Devetta}},
  \bibinfo{author}{\bibfnamefont{A.}~\bibnamefont{Bordoni}},
  \bibinfo{author}{\bibfnamefont{P.}~\bibnamefont{Piseri}}, \bibnamefont{and}
  \bibinfo{author}{\bibfnamefont{P.}~\bibnamefont{Milani}},
  \bibinfo{journal}{Phys. Rev. Lett} \textbf{\bibinfo{volume}{102}},
  \bibinfo{pages}{245502} (\bibinfo{year}{2009}).

\bibitem[{\citenamefont{Jin et~al.}(2009)\citenamefont{Jin, Lan, Peng, Suenaga,
  and Iijima}}]{Jin2009}
\bibinfo{author}{\bibfnamefont{C.}~\bibnamefont{Jin}},
  \bibinfo{author}{\bibfnamefont{H.}~\bibnamefont{Lan}},
  \bibinfo{author}{\bibfnamefont{L.}~\bibnamefont{Peng}},
  \bibinfo{author}{\bibfnamefont{K.}~\bibnamefont{Suenaga}}, \bibnamefont{and}
  \bibinfo{author}{\bibfnamefont{S.}~\bibnamefont{Iijima}},
  \bibinfo{journal}{Phys. Rev. Lett} \textbf{\bibinfo{volume}{102}},
  \bibinfo{pages}{205501} (\bibinfo{year}{2009}).

\bibitem[{\citenamefont{Casari et~al.}(2004)\citenamefont{Casari, Li~Bassi,
  Ravagnan, Siviero, Lenardi, Piseri, Bongiorno, Bottani, and Milani}}]{6}
\bibinfo{author}{\bibfnamefont{C.~S.} \bibnamefont{Casari}},
  \bibinfo{author}{\bibfnamefont{A.}~\bibnamefont{Li~Bassi}},
  \bibinfo{author}{\bibfnamefont{L.}~\bibnamefont{Ravagnan}},
  \bibinfo{author}{\bibfnamefont{F.}~\bibnamefont{Siviero}},
  \bibinfo{author}{\bibfnamefont{C.}~\bibnamefont{Lenardi}},
  \bibinfo{author}{\bibfnamefont{P.}~\bibnamefont{Piseri}},
  \bibinfo{author}{\bibfnamefont{G.}~\bibnamefont{Bongiorno}},
  \bibinfo{author}{\bibfnamefont{C.~E.} \bibnamefont{Bottani}},
  \bibnamefont{and} \bibinfo{author}{\bibfnamefont{P.}~\bibnamefont{Milani}},
  \bibinfo{journal}{Phys. Rev. B} \textbf{\bibinfo{volume}{69}},
  \bibinfo{pages}{075422} (\bibinfo{year}{2004}).

\bibitem[{\citenamefont{Magnano et~al.}(2003)\citenamefont{Magnano, Cepek,
  Sancrotti, Siviero, Vinati, Lenardi, Piseri, Barborini, and
  Milani}}]{Mag2003}
\bibinfo{author}{\bibfnamefont{E.}~\bibnamefont{Magnano}},
  \bibinfo{author}{\bibfnamefont{C.}~\bibnamefont{Cepek}},
  \bibinfo{author}{\bibfnamefont{M.}~\bibnamefont{Sancrotti}},
  \bibinfo{author}{\bibfnamefont{F.}~\bibnamefont{Siviero}},
  \bibinfo{author}{\bibfnamefont{S.}~\bibnamefont{Vinati}},
  \bibinfo{author}{\bibfnamefont{C.}~\bibnamefont{Lenardi}},
  \bibinfo{author}{\bibfnamefont{P.}~\bibnamefont{Piseri}},
  \bibinfo{author}{\bibfnamefont{E.}~\bibnamefont{Barborini}},
  \bibnamefont{and} \bibinfo{author}{\bibfnamefont{P.}~\bibnamefont{Milani}},
  \bibinfo{journal}{Phys. Rev. B} \textbf{\bibinfo{volume}{67}},
  \bibinfo{pages}{125414} (\bibinfo{year}{2003}).

\bibitem[{\citenamefont{Ravagnan et~al.}(2002)\citenamefont{Ravagnan, Siviero,
  Lenardi, Piseri, Barborini, Milani, Casari, Bassi, and Bottani}}]{Rav2002}
\bibinfo{author}{\bibfnamefont{L.}~\bibnamefont{Ravagnan}},
  \bibinfo{author}{\bibfnamefont{F.}~\bibnamefont{Siviero}},
  \bibinfo{author}{\bibfnamefont{C.}~\bibnamefont{Lenardi}},
  \bibinfo{author}{\bibfnamefont{P.}~\bibnamefont{Piseri}},
  \bibinfo{author}{\bibfnamefont{E.}~\bibnamefont{Barborini}},
  \bibinfo{author}{\bibfnamefont{P.}~\bibnamefont{Milani}},
  \bibinfo{author}{\bibfnamefont{C.~S.} \bibnamefont{Casari}},
  \bibinfo{author}{\bibfnamefont{A.~L.} \bibnamefont{Bassi}}, \bibnamefont{and}
  \bibinfo{author}{\bibfnamefont{C.~E.} \bibnamefont{Bottani}},
  \bibinfo{journal}{Phys. Rev. Lett} \textbf{\bibinfo{volume}{89}},
  \bibinfo{pages}{285506} (\bibinfo{year}{2002}).

\bibitem[{\citenamefont{Scemama et~al.}(2002)\citenamefont{Scemama, Chaquin,
  Gazeau, and Bénilan}}]{Sce2002}
\bibinfo{author}{\bibfnamefont{A.}~\bibnamefont{Scemama}},
  \bibinfo{author}{\bibfnamefont{P.}~\bibnamefont{Chaquin}},
  \bibinfo{author}{\bibfnamefont{M.-C.} \bibnamefont{Gazeau}},
  \bibnamefont{and} \bibinfo{author}{\bibfnamefont{Y.}~\bibnamefont{Bénilan}},
  \bibinfo{journal}{Chem. Phys. Lett} \textbf{\bibinfo{volume}{361}},
  \bibinfo{pages}{520} (\bibinfo{year}{2002}).

\bibitem[{\citenamefont{Hino et~al.}(2003)\citenamefont{Hino, Okada, Iwasaki,
  Kijima, and Shirakawa}}]{H04}
\bibinfo{author}{\bibfnamefont{S.}~\bibnamefont{Hino}},
  \bibinfo{author}{\bibfnamefont{Y.}~\bibnamefont{Okada}},
  \bibinfo{author}{\bibfnamefont{K.}~\bibnamefont{Iwasaki}},
  \bibinfo{author}{\bibfnamefont{M.}~\bibnamefont{Kijima}}, \bibnamefont{and}
  \bibinfo{author}{\bibfnamefont{H.}~\bibnamefont{Shirakawa}},
  \bibinfo{journal}{Chem. Phys. Lett} \textbf{\bibinfo{volume}{372}},
  \bibinfo{pages}{59} (\bibinfo{year}{2003}).

\bibitem[{\citenamefont{Grutter et~al.}(1998)\citenamefont{Grutter, Wyss,
  Fulara, and Maier}}]{4}
\bibinfo{author}{\bibfnamefont{M.}~\bibnamefont{Grutter}},
  \bibinfo{author}{\bibfnamefont{M.}~\bibnamefont{Wyss}},
  \bibinfo{author}{\bibfnamefont{J.}~\bibnamefont{Fulara}}, \bibnamefont{and}
  \bibinfo{author}{\bibfnamefont{J.}~\bibnamefont{Maier}}, \bibinfo{journal}{J.
  Phys. Chem. A} \textbf{\bibinfo{volume}{102}}, \bibinfo{pages}{9785}
  (\bibinfo{year}{1998}).

\bibitem[{\citenamefont{Ball et~al.}(2000)\citenamefont{Ball, McCarthy, and
  Taddheus}}]{5}
\bibinfo{author}{\bibfnamefont{C.}~\bibnamefont{Ball}},
  \bibinfo{author}{\bibfnamefont{M.}~\bibnamefont{McCarthy}}, \bibnamefont{and}
  \bibinfo{author}{\bibfnamefont{P.}~\bibnamefont{Taddheus}},
  \bibinfo{journal}{J. Chem. Phys.} \textbf{\bibinfo{volume}{112}},
  \bibinfo{pages}{10149} (\bibinfo{year}{2000}).

\bibitem[{\citenamefont{Hohenberg and Kohn}(1964)}]{Hoh64}
\bibinfo{author}{\bibfnamefont{P.}~\bibnamefont{Hohenberg}} \bibnamefont{and}
  \bibinfo{author}{\bibfnamefont{W.}~\bibnamefont{Kohn}},
  \bibinfo{journal}{Phys.~Rev.~} \textbf{\bibinfo{volume}{136 B}},
  \bibinfo{pages}{864} (\bibinfo{year}{1964}).

\bibitem[{\citenamefont{Kohn and Sham}(1965)}]{Koh65}
\bibinfo{author}{\bibfnamefont{W.}~\bibnamefont{Kohn}} \bibnamefont{and}
  \bibinfo{author}{\bibfnamefont{L.~J.} \bibnamefont{Sham}},
  \bibinfo{journal}{Phys.~Rev.~} \textbf{\bibinfo{volume}{140 A}},
  \bibinfo{pages}{1133} (\bibinfo{year}{1965}).

\bibitem[{\citenamefont{Ehrenreich and Cohen}(1959)}]{Ehr59}
\bibinfo{author}{\bibfnamefont{H.}~\bibnamefont{Ehrenreich}} \bibnamefont{and}
  \bibinfo{author}{\bibfnamefont{M.~H.} \bibnamefont{Cohen}},
  \bibinfo{journal}{Phys. Rev.} \textbf{\bibinfo{volume}{115}},
  \bibinfo{pages}{786} (\bibinfo{year}{1959}).

\bibitem[{\citenamefont{{\tt http://www.abinit.org}}()}]{ABINIT}
\bibinfo{author}{\bibnamefont{{\tt http://www.abinit.org}}}.

\bibitem[{\citenamefont{{\tt http://www.yambo-code.org}}()}]{YAMBO}
\bibinfo{author}{\bibnamefont{{\tt http://www.yambo-code.org}}}.

\bibitem[{\citenamefont{Hogan et~al.}(2003)\citenamefont{Hogan, DelSole, and
  Onida}}]{Hog03}
\bibinfo{author}{\bibfnamefont{C.}~\bibnamefont{Hogan}},
  \bibinfo{author}{\bibfnamefont{R.}~\bibnamefont{DelSole}}, \bibnamefont{and}
  \bibinfo{author}{\bibfnamefont{G.}~\bibnamefont{Onida}},
  \bibinfo{journal}{Phys.~Rev.~B} \textbf{\bibinfo{volume}{68}},
  \bibinfo{pages}{035405} (\bibinfo{year}{2003}).

\bibitem[{\citenamefont{Castillo et~al.}(2003)\citenamefont{Castillo, Mendoza,
  Schmidt, Hahn, and Bechstedt}}]{Cas03}
\bibinfo{author}{\bibfnamefont{C.}~\bibnamefont{Castillo}},
  \bibinfo{author}{\bibfnamefont{B.~S.} \bibnamefont{Mendoza}},
  \bibinfo{author}{\bibfnamefont{W.~G.} \bibnamefont{Schmidt}},
  \bibinfo{author}{\bibfnamefont{P.~H.} \bibnamefont{Hahn}}, \bibnamefont{and}
  \bibinfo{author}{\bibfnamefont{F.}~\bibnamefont{Bechstedt}},
  \bibinfo{journal}{Phys.~Rev.~B} \textbf{\bibinfo{volume}{68}},
  \bibinfo{pages}{041310(R)} (\bibinfo{year}{2003}).

\bibitem[{\citenamefont{Monachesi et~al.}(2003)\citenamefont{Monachesi,
  Palummo, DelSole, Grechnev, and Eriksson}}]{Mon03}
\bibinfo{author}{\bibfnamefont{P.}~\bibnamefont{Monachesi}},
  \bibinfo{author}{\bibfnamefont{M.}~\bibnamefont{Palummo}},
  \bibinfo{author}{\bibfnamefont{R.}~\bibnamefont{DelSole}},
  \bibinfo{author}{\bibfnamefont{A.}~\bibnamefont{Grechnev}}, \bibnamefont{and}
  \bibinfo{author}{\bibfnamefont{O.}~\bibnamefont{Eriksson}},
  \bibinfo{journal}{Phys.~Rev.~B} \textbf{\bibinfo{volume}{68}},
  \bibinfo{pages}{035426} (\bibinfo{year}{2003}).

\bibitem[{\citenamefont{Gonze et~al.}(2005)\citenamefont{Gonze, Riganese,
  Verstraete, Beuken, Pouillon, Caracas, Jollet, Torrent, Zerah, Mikami
  et~al.}}]{Gon05}
\bibinfo{author}{\bibfnamefont{X.}~\bibnamefont{Gonze}},
  \bibinfo{author}{\bibfnamefont{G.-M.} \bibnamefont{Riganese}},
  \bibinfo{author}{\bibfnamefont{M.}~\bibnamefont{Verstraete}},
  \bibinfo{author}{\bibfnamefont{J.-M.} \bibnamefont{Beuken}},
  \bibinfo{author}{\bibfnamefont{Y.}~\bibnamefont{Pouillon}},
  \bibinfo{author}{\bibfnamefont{R.}~\bibnamefont{Caracas}},
  \bibinfo{author}{\bibfnamefont{F.}~\bibnamefont{Jollet}},
  \bibinfo{author}{\bibfnamefont{M.}~\bibnamefont{Torrent}},
  \bibinfo{author}{\bibfnamefont{G.}~\bibnamefont{Zerah}},
  \bibinfo{author}{\bibfnamefont{M.}~\bibnamefont{Mikami}},
  \bibnamefont{et~al.}, \bibinfo{journal}{Z. Kristallogr.}
  \textbf{\bibinfo{volume}{220}}, \bibinfo{pages}{558} (\bibinfo{year}{2005}).

\bibitem[{\citenamefont{Ceperley and Alder}(1980)}]{Cep80}
\bibinfo{author}{\bibfnamefont{D.~M.} \bibnamefont{Ceperley}} \bibnamefont{and}
  \bibinfo{author}{\bibfnamefont{B.~J.} \bibnamefont{Alder}},
  \bibinfo{journal}{Phys.~Rev.~Lett.~} \textbf{\bibinfo{volume}{45}},
  \bibinfo{pages}{566} (\bibinfo{year}{1980}).

\bibitem[{\citenamefont{Perdew and Zunger}(1981)}]{Per81}
\bibinfo{author}{\bibfnamefont{J.~P.} \bibnamefont{Perdew}} \bibnamefont{and}
  \bibinfo{author}{\bibfnamefont{A.}~\bibnamefont{Zunger}},
  \bibinfo{journal}{Phys.~Rev.~B} \textbf{\bibinfo{volume}{23}},
  \bibinfo{pages}{5048} (\bibinfo{year}{1981}).

\bibitem[{\citenamefont{Troullier and Martins}(1991)}]{Tro91}
\bibinfo{author}{\bibfnamefont{N.}~\bibnamefont{Troullier}} \bibnamefont{and}
  \bibinfo{author}{\bibfnamefont{J.~L.} \bibnamefont{Martins}},
  \bibinfo{journal}{Phys.~Rev.~B} \textbf{\bibinfo{volume}{43}},
  \bibinfo{pages}{1993} (\bibinfo{year}{1991}).

\bibitem[{\citenamefont{Motta}(2007-2008)}]{Mot08}
\bibinfo{author}{\bibfnamefont{C.}~\bibnamefont{Motta}}, Master's thesis,
  \bibinfo{school}{Universit\'a degli Studi di Milano}
  (\bibinfo{year}{2007-2008}).

\bibitem[{\citenamefont{Marini et~al.}(2009)\citenamefont{Marini, Hogan,
  Gr{\"u}ning, and Varsano}}]{Mar2009}
\bibinfo{author}{\bibfnamefont{A.}~\bibnamefont{Marini}},
  \bibinfo{author}{\bibfnamefont{C.}~\bibnamefont{Hogan}},
  \bibinfo{author}{\bibfnamefont{M.}~\bibnamefont{Gr{\"u}ning}},
  \bibnamefont{and} \bibinfo{author}{\bibfnamefont{D.}~\bibnamefont{Varsano}},
  \bibinfo{journal}{Comp. Phys. Commun.} p.
  \bibinfo{pages}{doi:10.1016/j.cpc.2009.02.003} (\bibinfo{year}{2009}).

\bibitem[{\citenamefont{Onida et~al.}(2002)\citenamefont{Onida, Reining, and
  Rubio}}]{Oni02}
\bibinfo{author}{\bibfnamefont{G.}~\bibnamefont{Onida}},
  \bibinfo{author}{\bibfnamefont{L.}~\bibnamefont{Reining}}, \bibnamefont{and}
  \bibinfo{author}{\bibfnamefont{A.}~\bibnamefont{Rubio}},
  \bibinfo{journal}{Rev. Mod. Phys.} \textbf{\bibinfo{volume}{74}},
  \bibinfo{pages}{601} (\bibinfo{year}{2002}).

\bibitem[{\citenamefont{Bassani}(1975)}]{Bas75}
\bibinfo{author}{\bibfnamefont{G.~F.} \bibnamefont{Bassani}}, in
  \emph{\bibinfo{booktitle}{Electronic states and optical transitions in
  solids}} (\bibinfo{publisher}{Pergamon Press, Oxford}, \bibinfo{year}{1975}),
  p. \bibinfo{pages}{149}.

\bibitem[{\citenamefont{Motta et~al.}(2009)\citenamefont{Motta, Cazzaniga,
  Giantomassi, Ga{\'a}l-Nagy, Onida, and Gonze}}]{PREPRINT}
\bibinfo{author}{\bibfnamefont{C.}~\bibnamefont{Motta}},
  \bibinfo{author}{\bibfnamefont{M.}~\bibnamefont{Cazzaniga}},
  \bibinfo{author}{\bibfnamefont{M.}~\bibnamefont{Giantomassi}},
  \bibinfo{author}{\bibfnamefont{K.}~\bibnamefont{Ga{\'a}l-Nagy}},
  \bibinfo{author}{\bibfnamefont{G.}~\bibnamefont{Onida}}, \bibnamefont{and}
  \bibinfo{author}{\bibfnamefont{X.}~\bibnamefont{Gonze}},
  \bibinfo{journal}{Preprint}  (\bibinfo{year}{2009}).

\bibitem[{\citenamefont{Marques and Gross}(2004)}]{Mar2004}
\bibinfo{author}{\bibfnamefont{M.~A.~L.} \bibnamefont{Marques}}
  \bibnamefont{and} \bibinfo{author}{\bibfnamefont{E.~K.~U.}
  \bibnamefont{Gross}}, \bibinfo{journal}{Ann. Rev. Phys. Chem.}
  \textbf{\bibinfo{volume}{55}}, \bibinfo{pages}{427} (\bibinfo{year}{2004}).

\bibitem[{\citenamefont{Casida et~al.}(1998{\natexlab{a}})\citenamefont{Casida,
  Casida, and Salahub}}]{Cas1998a}
\bibinfo{author}{\bibfnamefont{M.~E.} \bibnamefont{Casida}},
  \bibinfo{author}{\bibfnamefont{K.~C.} \bibnamefont{Casida}},
  \bibnamefont{and} \bibinfo{author}{\bibfnamefont{D.~R.}
  \bibnamefont{Salahub}}, \bibinfo{journal}{Int. Jour. Quant. Chem.}
  \textbf{\bibinfo{volume}{70}}, \bibinfo{pages}{933}
  (\bibinfo{year}{1998}{\natexlab{a}}).

\bibitem[{\citenamefont{Casida et~al.}(1998{\natexlab{b}})\citenamefont{Casida,
  Jamorski, Casida, and Salahub}}]{Cas1998b}
\bibinfo{author}{\bibfnamefont{M.~E.} \bibnamefont{Casida}},
  \bibinfo{author}{\bibfnamefont{C.}~\bibnamefont{Jamorski}},
  \bibinfo{author}{\bibfnamefont{K.~C.} \bibnamefont{Casida}},
  \bibnamefont{and} \bibinfo{author}{\bibfnamefont{D.~R.}
  \bibnamefont{Salahub}}, \bibinfo{journal}{Jour. Chem. Phys.}
  \textbf{\bibinfo{volume}{108}}, \bibinfo{pages}{4439}
  (\bibinfo{year}{1998}{\natexlab{b}}).

\bibitem[{\citenamefont{Jamorski et~al.}(1996)\citenamefont{Jamorski, Casida,
  and Salahub}}]{Jam1996}
\bibinfo{author}{\bibfnamefont{C.}~\bibnamefont{Jamorski}},
  \bibinfo{author}{\bibfnamefont{M.~E.} \bibnamefont{Casida}},
  \bibnamefont{and} \bibinfo{author}{\bibfnamefont{D.~R.}
  \bibnamefont{Salahub}}, \bibinfo{journal}{Jour. Chem. Phys.}
  \textbf{\bibinfo{volume}{104}}, \bibinfo{pages}{5134} (\bibinfo{year}{1996}).

\bibitem[{\citenamefont{Casida and Wesolowski}(2004)}]{Cas2004}
\bibinfo{author}{\bibfnamefont{M.~E.} \bibnamefont{Casida}} \bibnamefont{and}
  \bibinfo{author}{\bibfnamefont{T.~A.} \bibnamefont{Wesolowski}},
  \bibinfo{journal}{Int. Jour. Quant. Chem.} \textbf{\bibinfo{volume}{96}},
  \bibinfo{pages}{577} (\bibinfo{year}{2004}).

\bibitem[{\citenamefont{Vasiliev et~al.}(1999)\citenamefont{Vasiliev, {\"O}gut,
  and Chelikowsky}}]{Vas1999}
\bibinfo{author}{\bibfnamefont{I.}~\bibnamefont{Vasiliev}},
  \bibinfo{author}{\bibfnamefont{S.}~\bibnamefont{{\"O}gut}}, \bibnamefont{and}
  \bibinfo{author}{\bibfnamefont{J.~R.} \bibnamefont{Chelikowsky}},
  \bibinfo{journal}{Phys. Rev. Lett.} \textbf{\bibinfo{volume}{82}},
  \bibinfo{pages}{1919} (\bibinfo{year}{1999}).

\end{thebibliography}

\end{document}